\documentclass[a4paper,preprint,5p,10pt]{elsarticle}
\usepackage{lipsum}
\usepackage{amssymb}
\usepackage{lineno,hyperref}
\modulolinenumbers[10]
\usepackage{tabularx}
\usepackage{float}
\biboptions{numbers,sort&compress}
\usepackage{mathrsfs}
\usepackage{graphicx}
    
\journal{Journal of Solid State Chemistry 10.1016/j.jssc.2018.12.064}

\begin{document}

\begin{frontmatter}

\title{Controlling the stoichiometry of the triangular lattice antiferromagnet Li$_{1+x}$Zn$_{2-y}$Mo$_3$O$_8$}

\author[add1]{Kim E. Sandvik
\corref{mycorrespondingauthor}}
\cortext[mycorrespondingauthor]{Corresponding author}
\ead{sandvikk@mail.tagen.tohoku.ac.jp}
\author[add1]{Daisuke Okuyama}
\author[add1]{Kazuhiro Nawa }
\author[add2,add3]{Maxim Avdeev}
\author[add1]{Taku J. Sato}

\address[add1]{Institute of Multidisciplinary Research for Advanced Materials, Tohoku University, Sendai 980-8577, Japan}
\address[add2]{Australian Nuclear Science and Technology Organization, Locked Bag 2001, Kirrawee DC NSW 2232, Australia}
\address[add3]{School of Chemistry, The University of Sydney, Sydney, NSW 2006, Australia}

\begin{abstract}
The control of the stoichiometry of Li$_{1+x}$Zn$_{2-y}$Mo$_3$O$_8$ was achieved by the solid-state-reaction. We found that the best sample that has the chemical composition Li$_{0.95(4)}$Zn$_{1.92(8)}$Mo$_3$O$_8$ was obtained from the starting nominal composition with Li\,:\,Zn\,:\,Mo\,:\,O = $(1+w)$\,:\,$(2.8-w)$\,:\,$3$\,:\,$8.6$ with $w = -0.1$, indicating that the stoichiometry is greatly improved compared to those in the earlier reports. For larger $w$ detailed structural analysis indicates that the mixed sites of Li and Zn are preferentially occupied by Li atoms, as well as the fraction of the non-magnetic secondary phase Zn$_2$Mo$_3$O$_8$ decreases. Magnetic susceptibility of the improved stoichiometry powder samples shows a broad hump in the temperature range of 100 $< T <$ 200\,K. This suggests that the development of antiferromagnetic correlations at the high temperatures is inherent to the ideal stoichiometric LiZn$_2$Mo$_3$O$_8$.
\end{abstract}


\end{frontmatter}

\section{Introduction}
\begin{NoHyper}\let\thefootnote\relax\footnote{\scriptsize{\textcopyright\,2019.\ This manuscript version is made available under the CC-BY-NC-ND 4.0 license http://creativecommons.org/licenses/by-nc-nd/4.0/}}\end{NoHyper}
Geometrically frustrated antiferromagnetic systems have been an intriguing topic in condensed matter physics for decades \cite{Wannier1950}. Lack of magnetic order is expected down to 0\,K in some of these systems where the formation of intriguing spin liquid or resonating valence bond state has been anticipated \cite{Balents2010}. Accordingly, there has been an increasing amount of experimental and theoretical activities in the study of frustrated 2D triangular \cite{Nakatsuji2005}, kagome \cite{Helton2007, Matan2010} and pyrochlore \cite{Reimers1991} lattice systems.

\begin{figure}[h]
\includegraphics[width=0.475 \textwidth]{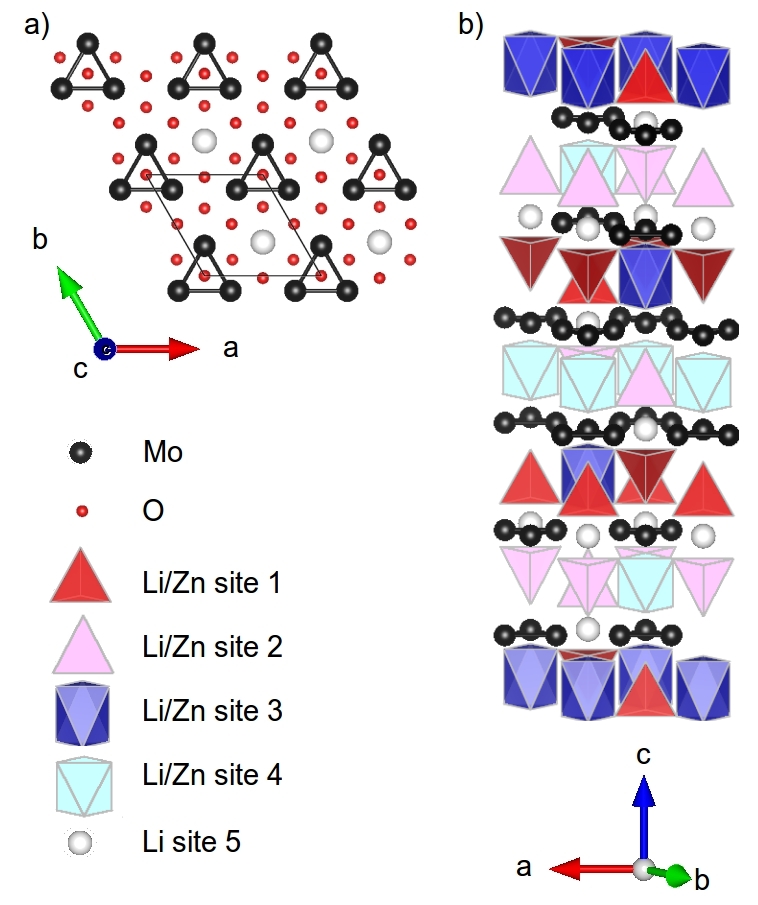}
\caption{\label{fig:ex1} Crystal structure of LiZn$_2$Mo$_3$O$_8$ (space group of $R\bar{3}m$, $a$ = 5.8\,$\textrm{\r{A}}$, $c$ = 31.1\,$\textrm{\r{A}}$). a) Illustration of the triangular lattice in the $ab$ plane. The thin black lines show the unit cell. Molybdenum sites in black are shown as clusters of three atoms aligned in $ab$ plane. Oxygen sites are shown as red spheres. b) The alternating stacking of the Mo$_3$O$_{8}$ triangular lattice layers and layers of Li/Zn atom sites along the $c$-axis. Li and Zn sites 1 -- 4 have intersite disorder. Tetrahedral sites 1 and 2 in red and magenta, respectively, tend to be Zn rich while octahedral sites 3 and 4 in blue and cyan, respectively, tend to be Li rich. Li site 5 is presented in white in the Mo planes.}
\label{fig:structure}
\end{figure}

Recently the study on the frustrated magnetism has expanded to cluster magnets. In these compounds the magnetic moment is delocalized over a group of atoms, called a cluster, instead of being localized on a single atom.
One group of such cluster magnets consists of the family of transition metal trimer compounds \cite{Haraguchi2015, Akbari-Sharbaf2018, Sheckelton2017, Haraguchi2017, Haraguchi2017_1, Torardi1985, Sheckelton2012, Flint2013, Mourigal2014, Sheckelton2014, Sheckelton2015, Chen2016}. Among them the two-dimensional triangular lattice antiferromagnet LiZn$_2$Mo$_3$O$_8$ has attracted considerable interest recently \cite{Sheckelton2012, Flint2013, Mourigal2014, Sheckelton2014, Sheckelton2015, Chen2016} since it was first reported in Ref. \cite{Torardi1985}. The compound has magnetic Mo$_3$O$_{13}$ clusters that form triangular lattices in the $ab$ plane (Fig. \ref{fig:structure}a). The triangular lattice planes are stacked along the $c$ direction and separated from each other by layers of Li$^+$ and Zn$^{2+}$ ions (Fig.\,\ref{fig:structure}b) which supply the planes with electrons. In the ideal case this will cause one unpaired spin 1/2 to be localized on each Mo$_3$O$_{13}$ cluster ([Mo$_3]^{11+}$).

Magnetic properties of LiZn$_2$Mo$_3$O$_8$ have been studied using various techniques, such as magnetic susceptibility \cite{Sheckelton2012, Mourigal2014, Sheckelton2015}, electron paramagnetic resonance \cite{Sheckelton2014}, and neutron inelastic scattering \cite{Mourigal2014}. To date all the studies indicate that the magnetic long-range order is absent down to $T$ = 0.05\,K in this compound. The temperature dependence of the magnetic susceptibility was analyzed in detail using Curie-Weiss fitting, indicating that there are two distinct temperature ranges with different effective moment sizes; for $T >$ 96\,K, the effective moment was approximately evaluated as 1.39\,$\mu_{\rm B}$, whereas $\sim$0.8\,$\mu_{\rm B}$ for the low temperature range 2 $< T <$ 96\,K.  From the reduction of the effective moment size around $\sim$96\,K, together with the absence of the long-range magnetic order, it was inferred that an intriguing condensed valence bond state is formed in this quantum triangular magnet at low temperatures \cite{Sheckelton2012}.

In LiZn$_2$Mo$_3$O$_8$, it has been known that there is chemical disorder in Li and Zn sites (see. Fig.\,\ref{fig:structure}b). This easily leads to off-stoichiometry. Indeed, neutron diffraction study indicates that the sample used in the earlier study suffers of this off-stoichiometry \cite{Sheckelton2012, Sheckelton2015}. It must be pointed out that the off-stoichiometry results in a hole doping of spin 1/2 electrons in the triangular lattice, so that the system cannot be regarded as the ideal spin 1/2 triangular antiferromagnet as it was originally expected. There is an attempt to electrochemically control the stoichiometry by removing Zn \cite{Sheckelton2015}. Nonetheless complete control of the stoichiometry including both Li and Zn concentrations has not been achieved as far as we are aware of. In this work, in order to achieve stoichiometry control in LiZn$_2$Mo$_3$O$_8$ we revisited the solid-state-reaction procedure, which was used in the original work, with widely changing starting compositions and heat treatment temperatures. We found a greatly improved condition which results in much better stoichiometry compared to the earlier work. The bulk magnetic properties of the obtained improved stoichiometry sample is investigated, which suggests that the formation of antiferromagnetic correlations at the higher temperatures 100 $< T <$ 200\,K is intrinsic to the triangular lattice physics of this material, rather than the lower temperature behavior.

\begin{figure}[h]
\includegraphics[width=0.475 \textwidth]{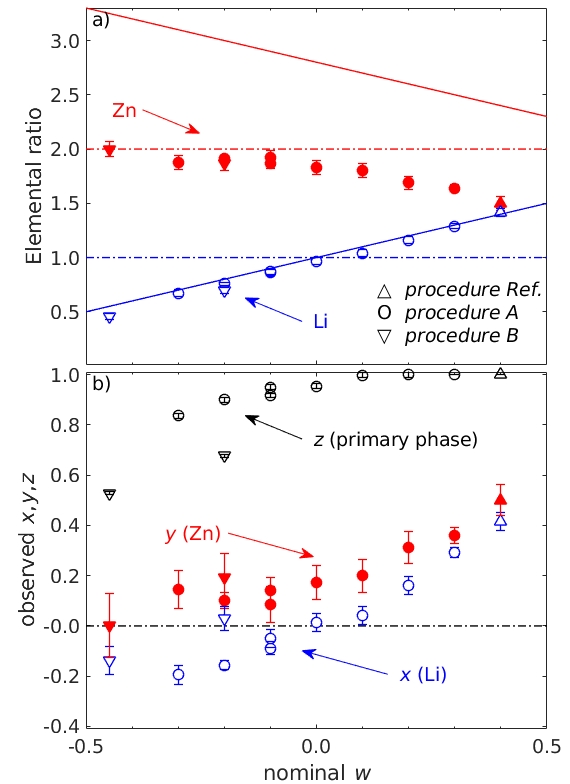}
\caption{\label{fig:ex1} Nominal starting composition $w$ dependence of various parameters. Up-pointing triangle, circle, and down-pointing triangle markers are associated with the $\it{procedure\ Ref}$, $\it{procedure\ A}$, and $\it{procedure\ B}$ samples, respectively. a) Elemental ratio of Li (Blue open markers) and Zn (red filled markers) obtained from ICP data in relation to postulated stoichiometric Mo concentration. Dash-dotted lines and solid lines are the stoichiometric ratio and nominal starting ratio, respectively. b) Nominal $w$ dependence of observed parameters $x$ (Li) (Blue open markers), $y$ (Zn) (red filled markers), and $z$ (primary phase) (black open markers) obtained by combined XRD + NPD Rietveld analysis. Black dash-dotted line shows the stoichiometric value for $x$ and $y$.}
\label{fig:wdependency}
\end{figure}

\section{Experimental}

Polycrystalline samples of Li$_{1+x}$Zn$_{2-y}$Mo$_3$O$_8$ were prepared by modifying a previously reported solid-state-reaction method \cite{Sheckelton2012}. The starting materials are Li$_2$MoO$_4$ (99$+\%$), ZnO (99.9$\%$), MoO$_2$ (99.9$+\%$), MoO$_3$ (99.9$+\%$) and Mo (99.9$\%$). Three different modifications in the solid-state-reaction procedure were tried in the present study. In the following they are referred to as $\it{procedure\ Ref}$ (almost the same procedure as reported in  Ref. \cite{Sheckelton2012}), $\it{procedure\ A}$, and $\it{procedure\ B}$. Assuming that the occupancy  of the disordered Li and Zn sites could be controlled by increasing Li$_2$MoO$_4$ while decreasing ZnO, we varied the nominal composition as Li\,:\,Zn\,:\,Mo\,:\,O = $(1+w)$\,:\,$(2.8-w)$\,:\,$3$\,:\,$(8.6+\sigma)$, where $w$ = 0.4 and $\sigma$ = 0 ($\it{procedure\ Ref}$), $-0.3 \leq w \leq 0.3$ and $\sigma = 0$ ($\it{procedure\ A}$) or $-0.45 \leq w \leq -0.2$ and $\sigma = 0.175$ ($\it{procedure\ B}$). A mixture of the Li$_2$MoO$_4$, ZnO, MoO$_2$, and Mo powders with the above molar ratio was used as the initial material for the solid-state-reaction for $\it{procedure\ Ref}$ and $\it{procedure\ B}$. For $\it{procedure\ A}$ MoO$_3$ is additionally used with a molar ratio of MoO$_2$:MoO$_3$ = 3:2 in addition to the Li$_2$MoO$_4$, ZnO, and Mo. 


%

\begin{figure}[h]
\includegraphics[width=0.51 \textwidth]{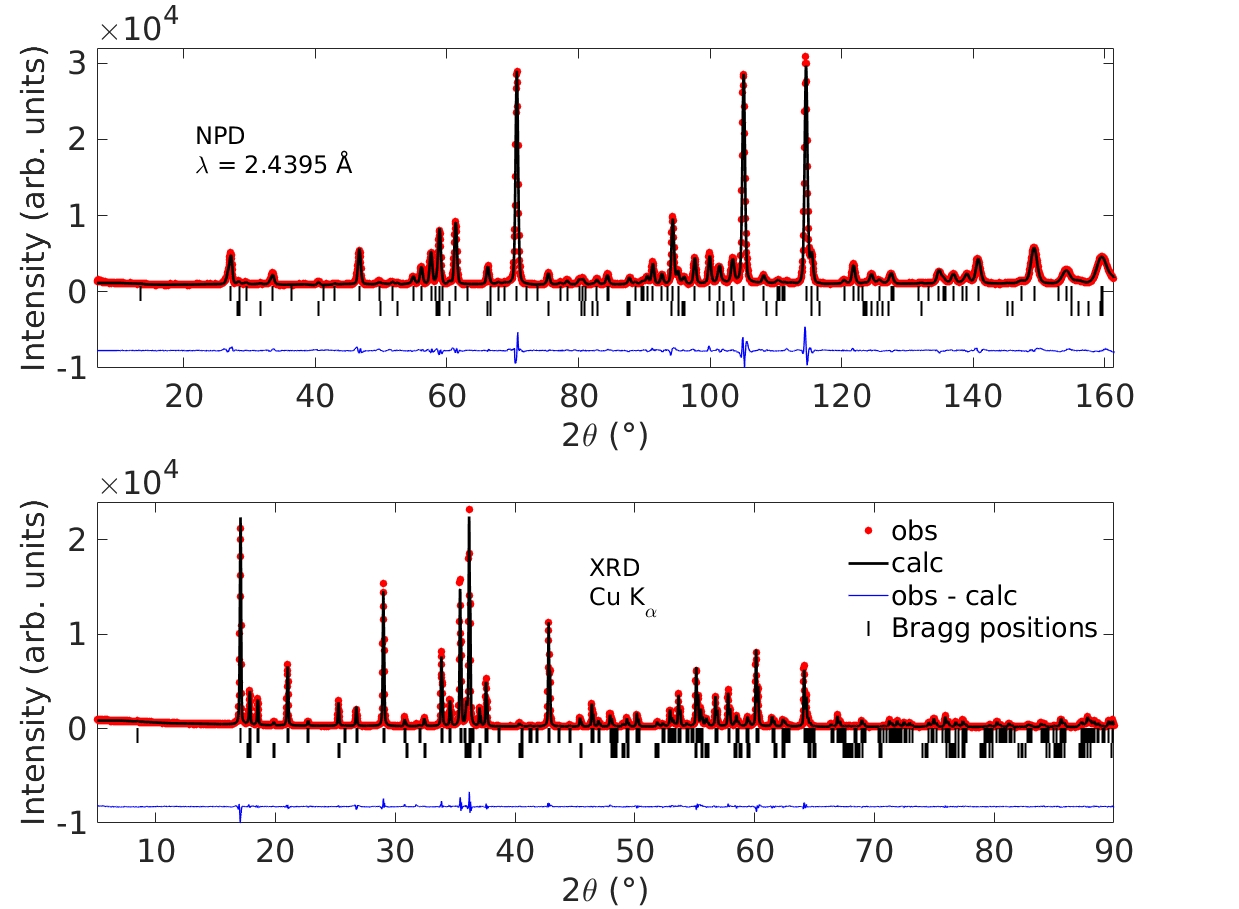}
\caption{\label{fig:ex1} Combined Rietveld analysis on Li$_{1+x}$Zn$_{2-y}$Mo$_3$O$_8$, $w = -0.1$ from high resolution powder diffraction at 300 K on ECHIDNA and x-ray diffraction patterns. The bottom blue lines give the difference between the observed (red dots) and calculated (black line) intensities. Bragg positions are shown as vertical bars in the upper and lower row for the primary and secondary phase, respectively.}
\label{fig:NPDXRD}
\end{figure}

Except for ZnO and Li$_2$MoO$_4$ which were dried at 160\,$^{\circ}$C before using, all the starting chemicals were used as received. The mixture of the chemicals was grinded, pelletized, and put into an Al$_2$O$_3$ crucible which was then evacuated and sealed in a quartz tube. For most of the solid-state-reactions the heat-treatment sequence started with a 100\,$^{\circ}$C/h ramp to 600\,$^{\circ}$C, followed by a 24\,h wait, before ramping up with 10\,$^{\circ}$C/h to the final temperature of 1050\,$^{\circ}$C. After keeping the final temperature for 12\,h, the reaction vessel was quenched into water. The final temperature of 1000\,$^{\circ}$C instead of 1050\,$^{\circ}$C was used for two samples: $\it{procedure\ A}$ with $w = 0$ and $\it{procedure\ Ref}$. 

A regrinding and second reaction sequence were done only for $\it{procedure\ Ref}$, otherwise only one reaction sequence was performed. 
As the final step, the reacted powder was washed with 3M HCl and rinsed with pure water several times to remove unreacted ingredients. Almost all $\it{procedure\ A}$ had a small and $\it{procedure\ B}$ a large amount of secondary phase which was identified as non-magnetic Zn$_2$Mo$_3$O$_8$. As a reference for the non-magnetic secondary phase, a polycrystalline sample of Zn$_2$Mo$_3$O$_8$ compound was also separately synthesized as previously reported in Ref. \cite{Sheckelton2012} and confirmed to be indeed non-magnetic by measuring it's magnetic susceptibility. 

Inductively coupled plasma (ICP) mass spectroscopy (Arcos EOP, Spectro) was used to determine the elemental ratios of Li, Zn, and Mo of the obtained powder samples. Three standard solutions with Li\,:\,Zn\,:\,Mo concentrations in ppm units as 0\,:\,0\,:\,0,\,\,\,0.508\,:\,10.04\,:\,20.06 and 1.016\,:\,20.08\,:\,40.12 were measured so that the data could be evaluated on a calibration curve using the variance-covariance matrix error analysis \cite{calibration_curve}. Assuming Mo to be stoichiometric, relative amounts of Li and Zn were calculated. For the structural characterization of the obtained powders, we used two diffraction techniques having a different beam source. One is the x-ray powder diffraction performed using the Cu K$_\alpha$ radiation (Ultima IV, Rigaku) and scanned in the range of 5 $\leq$ 2$\theta$ $\leq$ 90$^{\circ}$ in steps of 0.02$^{\circ}$ in room temperature. The other is the neutron powder diffraction (NPD) performed using the high resolution powder diffractometer ECHIDNA at the OPAL research reactor at ANSTO. The angular range of 6.5 $\leq$ 2$\theta$ $\leq$ 164$^{\circ}$ was scanned in steps of 0.05$^{\circ}$ at room temperature. Neutrons with the wavelength of 2.4395(5)\,$\textrm{\r{A}}$ were selected using the (331) reflections of the Ge monochromator. Utilizing the difference in the scattering lengths for the neutrons and x-rays, we performed combined XRD + NPD Rietveld analysis using the Fullprof software \cite{EDPCR} to obtain the Li and Zn compositions in the primary phase, as well as to determine the fraction of the secondary phase. The ICP results were used to constrain the total Li and Zn composition in both the primary and secondary phase when performing the combined NPD + XRD Rietveld analysis. Initial refinement parameters were taken from \citep{Sheckelton2015} and \cite{ZMO_structure} for the primary and secondary phase, respectively. From the refined parameters, the chemical composition of the primary phase, as well as its phase fraction in each powder sample, were estimated. The result is given in terms of three parameters $x,y$, and $z$ with which the chemical formula of the powder sample is expressed as $z$(Li$_{1+x}$Zn$_{2-y}$Mo$_3$O$_8$) $+$ ($1-z$)(Zn$_2$Mo$_3$O$_8$). The details of the refinement parameters, including number of free parameters, are summarized in Table. \ref{table:samples}.

Superconducting quantum interference device (SQUID) magnetometer (MPMS-XL, Quantum Design) was used to measure the magnetic susceptibility in the range of 2 $< T <$ 300\,K under 1\,T magnetic field. We describe the observed magnetization as $M_{\rm{obs}}/B$ = $\chi_{\rm{cell}}$ + $m_{\rm{tot}}[z\chi\rm_{1st}/\mathcal{M}_{1st}$ + $(1-z)\chi\rm_{2nd}/\mathcal{M}_{2nd}]/\it SF$, where $B$, $\chi_{\rm{cell}}$, $m_{\rm{tot}}$, $\chi\rm_{1st}$, $\chi\rm_{2nd}$, $\mathcal{M}_{1st}$, $\mathcal{M}_{2nd}$, and $SF$ are the magnetic field, magnetic susceptibility of the sample cell, total mass of the sample, magnetic susceptibilities of the primary and secondary phases, molar masses of the primary and secondary phases, and sample shape factor \cite{MPMS_effects}, respectively. After obtaining $\chi\rm_{1st}$ from the observed $M_{\rm{obs}}$, we estimate the intrinsic magnetic susceptibility of the primary phase as $\chi$ = $\chi\rm_{1st} - \chi\rm_0$, where $\chi\rm_0 = -3.68(1)\times10^{-5}$\,emu is the diamagnetic contribution which we approximate to be the same as for the non-magnetic secondary phase Zn$_2$Mo$_3$O$_8$ ($\chi_0 \sim \chi\rm_{2nd}$). By assuming no correlations between the variable errors, the variance of $\chi$ and similarly $\chi^{-1}$ is estimated as $\sigma_{\chi}^2 = \Sigma_i^n |\partial f(u_1,u_2,...,u_n)/\partial u_i|^2 \sigma_{u_i}^2$, where $\{u_1,u_2,...,u_n\}$ = $\{z, m_{\rm{tot}\it}, M_{\rm{obs}\it} \chi_{\rm{cell}\it}, \chi\rm_{1st}\it, \chi\rm_{2nd}\it, \chi\rm_{0}\it, \mathcal{M}_{1st}, \mathcal{M}_{2nd}, SF\}$.

\begin{table*}[h]
\scriptsize
\caption{\label{tab:ex1} Crystallographic data of Li$_{1+x}$Zn$_{2-y}$Mo$_3$O$_8$ at 300 K from the combined NPD + XRD Rietveld refinement with the ICP data being used as constrains when determing $x$ and $y$. Chemical formula unit $Z$ is 6 as well as space group is $R\overline{3}m$ (No. 166). The ratio of the primary phase is $z$ while the secondary impurity phase Zn$_2$Mo$_3$O$_8$ is ($1-z$). Unpaired spin 1/2 concentration per Mo$_3$O$_{13}$ cluster is given as $1 - 2y + x$. Lattice constants and unit cell volume are $a$, $c$, and $V$. Isotropic atomic displacement parameters are $B_{\rm{Mo}}$, $B_{\rm{O}}$, and $B_{\rm{Li/Zn}}$ where the subscript indicates the corresponding atom. Variables is the number of free parameters used for the Rietveld refinement in its final iteration. NPDs and XRDs are the number of Bragg-reflections in the x-ray and neutron diffraction data that were included in the Rietveld refinement. The agreement between observed and calculated peaks are expressed as $R_p$, $R_{wp}$, $R_{exp}$, and $\chi^2$ that are the profile factor, weighted profile factor, expected weighted profile factor, and Chi-squared, respectively.}
\label{table:samples}
\resizebox{\textwidth}{!}{\begin{tabular}{lccccccccccc}
\hline \hline
$\it{procedure}$ & $\it{Ref}$ & $\it{A}$ & $\it{A}$ & $\it{A}$ & $\it{A}$ & $\it{A}$ & $\it{A}$ & $\it{A}$ & $\it{A}$ & $\it{B}$ & $\it{B}$  \\
$w$ & 0.4 & 0.3 & 0.2 & 0.1 & 0 & $-$0.1& $-$0.1 & $-$0.2 & $-$0.3 & $-$0.2 & $-$0.45 \\
\hline
$x$ & 0.42(4) & 0.29(2) & 0.16(4) & 0.04(4) & 0.02(4) & $-$0.09(3)& $-$0.05(4) & $-$0.16(2) & $-$0.19(4) & 0.03(5) & 0.14(6)  \\
$y$ & 0.50(7) & 0.36(4) & 0.31(7) & 0.20(7) & 0.17(7) & 0.14(5)& 0.08(8) & 0.10(4) & 0.14(8) & 0.19(10) & 0.00(13)  \\
$z$ & 1 & 1 & 1.000(11) & 0.998(11) & 0.953(11) & 0.949(10)& 0.916(9) & 0.899(9) & 0.836(10) & 0.676(8) & 0.528(7)  \\
$1-2y+x$ & 0.42(13) & 0.57(7) & 0.54(14) & 0.64(14) & 0.67(15) & 0.63(11)& 0.78(15) & 0.64(7) & 0.52(16) & 0.65(20) & 0.86(26)  \\
$a$ ($\textrm{\r{A}}$) & 5.7843(3) & 5.7911(3) & 5.7935(3) & 5.7968(3) & 5.7968(3) & 5.8016(3)& 5.8016(3) & 5.8053(3) & 5.8056(3) & 5.7968(3) & 5.8052(3)  \\
$c$ ($\textrm{\r{A}}$) & 31.053(2) & 31.082(2) & 31.081(2) & 31.089(2) & 31.082(2) & 31.093(2)& 31.094(2) & 31.101(2) & 31.104(2) & 31.091(2) & 31.097(2)  \\
$V$ ($\textrm{\r{A}}^3$) & 899.78(9) & 902.72(9) & 903.48(9) & 904.72(9) & 904.52(9) & 906.33(9)& 906.36(9) & 907.72(9) & 907.91(9) & 904.76(9) & 907.58(9)  \\
$B_{\rm{Mo}}$ ($\textrm{\r{A}}^2$) & 0.42(3) & 0.51(3) & 0.80(3) & 0.67(3) & 0.53(3) & 0.68(3)& 0.48(3) & 0.58(3) & 0.63(3) & 0.54(3) & 0.57(4)  \\
$B_{\rm{O}}$ ($\textrm{\r{A}}^2$) & 0.48(4) & 0.45(4) & 0.57(4) & 0.58(4) & 0.49(4) & 0.60(4)& 0.45(3) & 0.61(4) & 0.52(4) & 0.48(5) & 0.57(5)  \\
$B_{\rm{Li/Zn}}$ ($\textrm{\r{A}}^2$) & 0.59(7) & 0.57(6) & 0.56(6) & 0.60(5) & 0.54(6) & 0.57(6)& 0.46(5) & 0.58(5) & 0.56(6) & 0.47(6) & 0.54(7)  \\
Variables & 27 & 27 & 30 & 30 & 40 & 40& 40 & 40 & 40 & 43 & 43  \\
NPDs & 88 & 88 & 134 & 134 & 134 & 134& 134 & 134 & 134 & 134 & 134  \\
XRDs & 257 & 250 & 387 & 386 & 389 & 389& 388 & 389 & 389 & 389 & 391  \\
$R_{\rm{p}}$ (\%) & 5.89 & 5.51 & 5.4 & 5.41 & 5.39 & 5.24& 4.96 & 5.22 & 5.74 & 4.62 & 4.52  \\
$R_{\rm{wp}}$ (\%) & 7.76 & 7.09 & 7.17 & 7.02 & 7.02 & 6.81& 6.43 & 6.68 & 7.34 & 6.05 & 6.05  \\
$R_{\rm{exp}}$ (\%) & 3.39 & 4.02 & 3.9 & 4.01 & 3.9 & 3.24& 3.69 & 3.82 & 4.05 & 3.86 & 3.73  \\
$\chi^2$ & 5.34 & 3.17 & 3.37 & 3.1 & 3.31 & 4.59& 3.05 & 3.1 & 3.32 & 2.52 & 2.65  \\
\hline \hline
\end{tabular}}
\end{table*}

\begin{table}[t]
\scriptsize
\caption{\label{tab:ex2} Fractional coordinates and site occupancies for Li$_{1+x}$Zn$_{2-y}$Mo$_3$O$_8$, $w = -0.1$, $x = -0.05(4)$, $y = 0.08(8)$, $z = 0.916(9)$. Isotropic atomic displacement parameters are $B_\textup{Mo} = 0.48(3)$, $B_\textup{O} = 0.45(3)$, and $B_\textup{Li/Zn} = 0.46(5)\,\textrm{\r{A}}^2$.}
\label{table:Am01_2}
\resizebox{\columnwidth}{!}{\begin{tabular}{lccccc}
\hline \hline
Atom, $i$  & Site & $a_i/a$ & $b_i/b$ & $c_i/c$ & occupancy \\
\hline
Mo1 & 18h & 0.18493(8) & 0.81507(8) & 0.08372(5) & 1  \\
O1 & 18h & 0.8451(3) & 0.1549(3) & 0.04788(12) & 1  \\
O2 & 18h & 0.4924(3) & 0.5076(3) & 0.12458(12) & 1  \\
O3 & 6c & 0 & 0 & 0.11841(18) & 1  \\
O4 & 6c & 0 & 0 & 0.37140(18) & 1  \\
Zn1 & 6c & 1/3 & 2/3 & $-$0.64229(9) & 0.933(4)  \\
Li1 & 6c & 1/3 & 2/3 & $-$0.64229(9) & $-$0.00(3)  \\
Zn2 & 6c & 0 & 0 & 0.18144(10) & 0.759(4)  \\
Li2 & 6c & 0 & 0 & 0.18144(10) & 0.25(3)  \\
Zn3 & 3a & 0 & 0 & 0 & 0.323(6)  \\
Li3 & 3a & 0 & 0 & 0 & 0.48(4)  \\
Zn4 & 6c & 0 & 0 & 0.5051(15) & 0.063(3)  \\
Li4 & 6c & 0 & 0 & 0.5051(15) & 0.48(2)  \\
Li5 & 6c & 2/3 & 1/3 & 0.08392 & $-$0.003(16)  \\
\hline \hline
\end{tabular}}
\end{table}

\section{Results and discussion}
First, chemical compositions of all the obtained powder samples were checked by the ICP mass spectroscopy. Elemental compositions for Li and Zn are summarized in Table. \ref{table:samples}. It may be noted that the composition obtained in the ICP analysis is the weighted average of the primary and secondary phase compositions, i.e. $z(1+x)$ and $(2-zy)$ for Li and Zn. To investigate the initial composition-parameter ($w$) dependence, the elemental compositions are plotted as a function of $w$ in Fig.\,\ref{fig:wdependency}a. Both the Li and Zn compositions show linear $w$ dependence of opposite sign. Zn does not surpass the stoichiometric value (dash-dotted line) while Li does so. In fact, the Li composition in the final product is almost the same as that of the starting material (solid line), while there is a large decrease in the Zn composition from the starting material to the final product, as unreacted ZnO has been removed by the HCl wash. It may be noted that the Zn composition starts to decrease when Li surpasses the stoichiometric line. This suggests that excess Li may replace the Zn atoms at the Li/Zn mixed sites.

Next, to determine the phase fraction of the primary Li$_{1+x}$Zn$_{2-y}$Mo$_3$O$_8$ and secondary Zn$_2$Mo$_3$O$_8$ phases, as well as to determine the site occupancies in the primary phase, we have performed the XRD and NPD experiments. Representative XRD and NPD patterns for the sample with $w = -0.1$ prepared using the $\it{procedure\ A}$ are given in Fig.\,\ref{fig:NPDXRD}. The XRD and NPD patterns are then simultaneously analyzed using the Rietveld technique. The result of the simultaneous Rietveld fitting is also shown in Fig.\,\ref{fig:NPDXRD}. Both the XRD and NPD patterns are satisfactorily reproduced by the Rietveld fitting, confirming the validity of the obtained crystallographic parameters. Fractional coordinates and occupancy parameters for the representative datasets are summarized in Table.$\:$\ref{table:Am01_2}. From these crystallographic parameters, we obtain $x = -0.05(4)$, $y = 0.08(8)$, and $z = 0.916(9)$ for the $w = -0.1$ powder sample. Applying the same analysis, we obtain the chemical composition parameters for all the other powder samples prepared from different initial composition with different procedures. The resulting chemical compositions, together with main crystallographic parameters, are shown in Table \ref{table:samples}. In the table, the experimental uncertainty given in the parentheses for $x$ and $y$ are derived from the ICP results, whereas those for the other parameters are from the Rietveld refinement. The $R$-factors and $\chi^2$ are reasonably small for all the refinements, indicating that the chemical compositions are accurately obtained in the present analysis.

\begin{figure}[h]
\includegraphics[width=0.475 \textwidth]{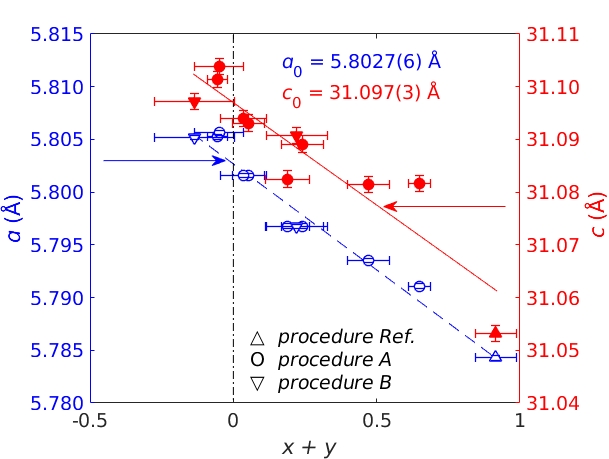}
\caption{\label{fig:ex1} Dependence of lattice constants $a$ and $c$ on $x+y$ in open blue markers (left axis) and filled red markers (right axis), respectively. The black dash-dotted line shows the stoichiometric value of $x + y = 0$. Up-pointed triangle, circle and down-pointed triangle markers are associated with the $\it{procedure\ Ref}$, $\it{procedure\ A}$, and $\it{procedure\ B}$ samples, respectively.}
\label{fig:vegard}
\end{figure}

To visualize the $w$-dependence, the parameters $x$, $y$, and $z$ are also plotted as a function of $w$ in Fig.\,\ref{fig:wdependency}b. The primary-phase-fraction $z$ becomes almost unity for $w$ $>$ 0, whereas the fraction of the secondary phase significantly increases as $w$ becomes smaller than 0. The Li concentration parameter $x$ almost monotonically increases as a function of $w$, with the stoichiometry concentration realized at $w$ $\sim$0. On the other hand, the Zn concentration parameter $y$ is mostly $w$-independent for $w$ $<$ 0, whereas it shows significant increase as $w$ becomes larger than 0. From those $w$-dependencies, we conclude that the best sample closest to the stoichiometry is obtained for the starting nominal composition parameter $w = -0.1$ with the preparation $\it{procedure\ A}$, which was indeed chosen as ``representative'' in the earlier paragraph. As noted before, the composition and phase-fraction parameters are  $x = -0.05(4)$, $y = 0.08(8)$, and $z = 0.916(9)$ for the $w = -0.1$. It may be noteworthy that they correspond to the unpaired spin 1/2 concentration per Mo$_3$O$_{13}$ cluster as $1- 2y + x$ = 0.78(15), which is greatly improved from 0.42(13) of the $\it{procedure\ Ref}$ sample obtained using exactly the same manner as the earlier report \cite{Sheckelton2012}.

In Fig.\,\ref{fig:vegard} the lattice constants $a$ and $c$ are plotted against $x + y$, which is the sum of the positive deviation of Li and negative deviation of Zn from the stoichiometry. This suggests that the replacement of Zn with Li results in decrease of the lattice constants obeying the Vegard's law. Knowing that both the lattice constants dominantly depend on the total deviation $x$ + $y$, we fit ($x + y$)-dependence of $a$ and $c$ using a linear function.  The fitting results are also shown in Fig.\,\ref{fig:vegard}. The lattice constants $a$ and $c$ are found to be linearly approximated to the following functions: $a$ = $-0.020(2)(x+y)$ + $a_0$ and $c$ = $-0.039(6)(x+y)$ + $c_0$, with the estimated lattice constants for the stoichiometric compound $a_0$ = 5.8027(6)$\,\textrm{\r{A}}$ and $c_0$ = 31.097(3)$\,\textrm{\r{A}}$. 

\begin{figure}[h]
\includegraphics[width=0.49 \textwidth]{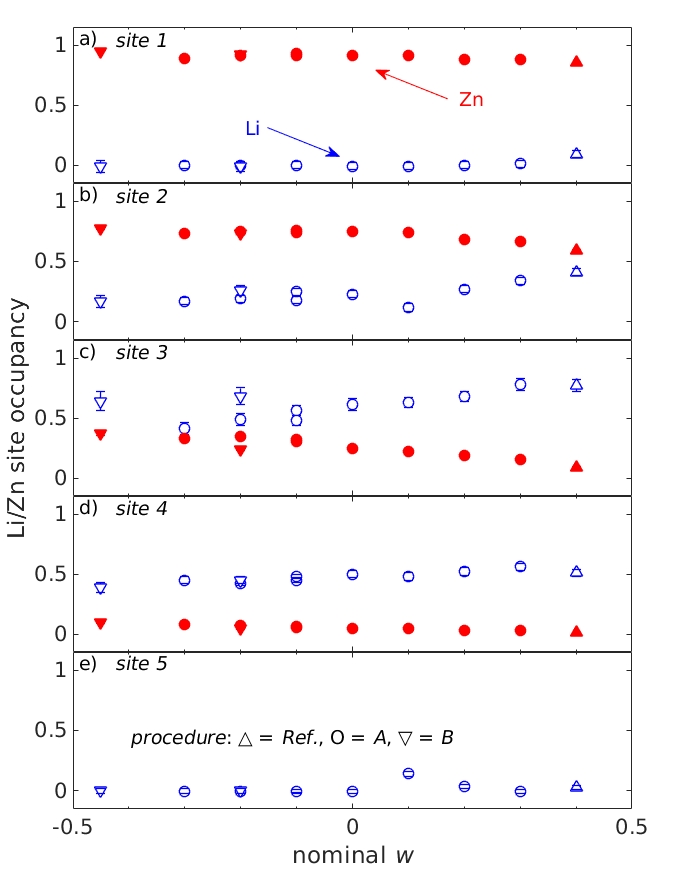}
\caption{\label{fig:ex1} $w$ dependence of the occupation of Li/Zn sites. Panels a) -- e) represent $\it{Sites\,1}$ -- $\it{5}$, respectively. Up-pointed triangle, circle, and down-pointed triangle markers are associated with the $\it{procedure\ Ref}$, $\it{procedure\ A}$, and $\it{procedure\ B}$ samples, respectively.}
\label{fig:site_occupation}
\end{figure}

Let us compare the stoichiometry of the previously reported sample \cite{Sheckelton2015} with our samples. For the previously reported sample without doping, the total deviation $x + y$ is estimated as 0.2(2) \cite{Sheckelton2015}. In addition, using the fit in Fig.\,\ref{fig:vegard} the lattice constants of a = 5.80163(3)$\,\textrm{\r{A}}$ and c = 31.0738(2)$\,\textrm{\r{A}}$ \cite{Sheckelton2015}  indicate $x + y$ to be 0.05(3) and 0.60(11), respectively. Thus, the stoichiometry of the previously reported sample should be not so poor as that of the $\it{procedure\ Ref}$, but still worse than that of the $w$ = $-$0.1 sample. This comparison also indicates that the stoichiometry is likely to be sample-dependent even if synthesis is performed in the same procedure (see Table 2. in the supplementary material for a more complete comparison). 

To investigate the site dependence of the Li and Zn composition variation, the $w$ dependent occupancy of the four Li/Zn sites and fifth Li site is presented in Fig.\,\ref{fig:site_occupation}. For the Zn rich tetrahedral sites, $\it{site\,1}$ is almost only occupied by Zn, whereas $\it{site\,2}$ have large fluctuation in Li occupation ranging as 0.12 -- 0.41. Similarly for the Li rich octahedral sites, there is only a small Zn occupation 0.012 -- 0.094 in $\it{site\,4}$ compared to the largely varying Zn occupation 0.087 -- 0.367 in $\it{site\,3}$. $\it{Site\,5}$ has zero occupancy of Li except for the $w$ = 0.1, 0.2, and 0.4 samples where tiny inclusion of Li improved the refinement slightly. The site preferences of Li and Zn atoms are consistent with those of previous samples \cite{Sheckelton2012,Sheckelton2015}.

\begin{figure}[h]
\includegraphics[width=0.475 \textwidth]{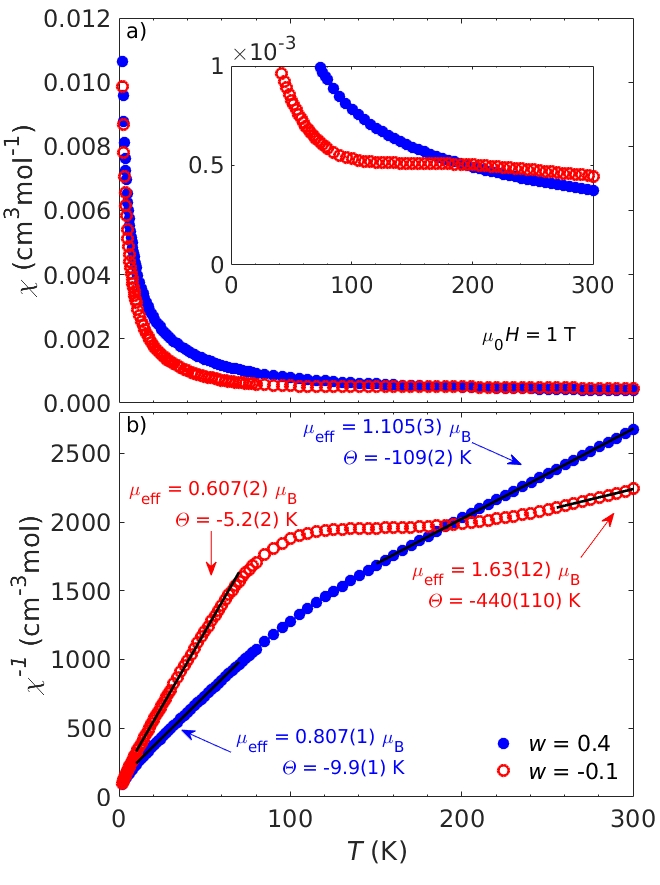}
\caption{\label{fig:ex1} Magnetic susceptibility $\chi$ (a) and inverse magnetic susceptibility $\chi^{-1}$ (b) of $w$ = 0.4 ($\it{procedure\ Ref}$) and $w = -0.1$ ($\it{procedure\ A}$), as blue filled circles and red open circles, respectively. The inset shows the data in narrower range of $\chi$ in order to emphasize the broad hump in sample $w = -0.1$. Linear fits on $\chi^{-1}$ are made to obtain the Curie constants $C$ and Weiss temperatures $\Theta$ from the low (10 $< T <$ 70\,K) and high (150 $< T <$ 300\,K for $w$ = 0.4 or 255 $< T <$ 300\,K for $w = -0.1$) temperature regions.}
\label{fig:susceptibility}
\end{figure}

Finally, we have performed magnetic susceptibility measurements for all the obtained samples with different preparation conditions to elucidate the initial composition dependence of the magnetic properties, and to conjecture the intrinsic magnetism of the stoichiometric sample. The representative results for the temperature dependence of the magnetic susceptibility are shown in Fig.\,\ref{fig:susceptibility}a for the best stoichiometry sample with $w = -0.1$ prepared using

\noindent $\it{procedure\ A}$, and for the reference one with $w = 0.4$ from $\it{procedure\ Ref}$. Corresponding inverse susceptibility is shown in Fig.\,\,\ref{fig:susceptibility}b. Data for all the other samples are given in the supplementary material. For 0.2 $\leq w \leq$ 0.4, where primary phase shows relatively large deviation from the stoichiometry, inverse susceptibility shows distinct linear behaviors for two temperature ranges; $T$ $>$ 100\,K and $T$ $<$ 100\,K (Suppl. Fig. 1). The appearance of the two distinct slopes in the Curie-Weiss behavior is in good agreement with the results reported in the earlier study \cite{Sheckelton2012}. The Curie-Weiss fitting was performed for the two temperature ranges separately using weighted linear regression with $\chi^{-1}$ = ($T-\Theta$)/$C$ as the model function where the variance of $\chi^{-1}$ is used for the weight. From the Curie constant $C$, we found that the effective moment size for 150 $< T <$ 300\,K is 1.105(3)\,$\mu_{\rm B}$, whereas it becomes greatly reduced at lower temperature as 0.807(1)\,$\mu_{\rm B}$ for 10 $< T <$ 70\,K for the $w$ = 0.4 sample. This is also quantitatively consistent with the earlier results \cite{Sheckelton2012,Sheckelton2015}. For the samples with better stoichiometry $w$ $<$ 0.2, however, a prominent feature was observed in the present study. As shown in Fig.\,\ref{fig:susceptibility}a, the magnetic susceptibility exhibits a weak hump around $\sim$160\,K, as exemplified by the sample with $w = -0.1$. Concomitantly, the Weiss temperature and effective moment size evaluated from the lower temperature Curie-Weiss fitting shows significant reduction as the system gets closer to the stoichiometry. From those observations, we conjecture that in the stoichiometric sample some sort of antiferromagnetic correlations is developed around 160\,K, which is also consistent with very large negative Weiss temperature. We note here that improving stoichiometry brings the low-temperature effective moment size (0.607(2)\,$\mu_{\rm B}$) further away from the value of 0.95\,$\mu_{\rm B}$ ($g = 1.9$) that is of the one-third of remaining paramagnetic spins. This strongly suggests that the remaining Curie-Weiss behavior in the low-temperature range with weaker antiferromagnetic correlations may be due to the remaining spins originating from the excess holes. We, hence, postulate that the inherent characteristic of the stoichiometric quantum triangular lattice antiferromagnet LiZn$_2$Mo$_3$O$_8$ is the antiferromagnetic correlation (i.e. formation of singlet ground state) dominated by exchange interactions of a few hundred Kelvin. The diminishing behavior of the remaining spins with improving stoichiometry further casts some doubt on the idea of the condensed valence bond ground state proposed in the previous study. Details of the putative antiferromagnetic correlations developing at high temperature, as well as origin of the remaining magnetic moments at lower temperatures are not clear at the present moment, and further microscopic studies using neutron scattering and/or electron paramagnetic resonance on better stoichiometry sample is highly desired.

\section{Conclusions}

Solid-state-reaction procedure was revisited to achieve stoichiometry control of the LiZn$_2$Mo$_3$O$_8$ compound.
By fine-tuning the initial nominal composition and varying selection of starting compounds, we found that the powder sample with much better stoichiometry can be obtained as Li$_{0.95(4)}$Zn$_{1.92(8)}$Mo$_3$O$_8$. The unpaired spin 1/2 concentration becomes much closer to 1 as 0.78(15) from 0.42(13) of the reference sample prepared in the same manner as the earlier works. From the magnetic susceptibility measurements, we found that in the improved stoichiometry sample antiferromagnetic correlations or an intriguing non-magnetic ground state forms below $\sim$160\,K, suggesting that this is the intrinsic magnetic properties of this quantum spin 1/2 triangular cluster antiferromagnet.

\section{Acknowledgments}
We thank M. Hino and T. Akiyama for ICP spectroscopy and A. P. Tsai for x-ray diffraction. This work was partly supported by Grants-in-Aid for Scientific Research (JSPS KAKENHI) (grant No. 17K18744). Work at IMRAM
is partly supported by the research program ``Dynamic alliance for open innovation bridging human, environment, and materials''. Travel expenses for the experiment at ECHIDNA was partly sponsored by the General User Program of ISSP-NSL, University of Tokyo.


\bibliography{LZMO_paper_Sandvik}

\end{document}